\documentclass[twocolumn,amsmath,amssymb,pre]{revtex4-1}

\usepackage{graphicx}% Include figure files
\usepackage{dcolumn}% Align table columns on decimal point
\usepackage{natbib}
\usepackage{amsmath,amsthm}
\usepackage{amssymb}
\usepackage{epstopdf}
\usepackage{times}
\usepackage{mathrsfs}
\usepackage{color}
\usepackage{gensymb}
\usepackage{times}
\usepackage{subfigure}
\usepackage[]{graphicx}
\usepackage{amssymb}
\usepackage{amsmath}
\usepackage{bm}
\DeclareGraphicsExtensions{.PNG,.jpg,.pdf,.gif,.eps}
\usepackage{amsmath,amssymb,amsthm,amsfonts,eucal}
\usepackage[english]{babel}

\begin{document}
\title{An asymmetry between pushing and pulling for crawling cells}
\author{Pierre Recho and Lev Truskinovsky}
%\email{}

\affiliation{ LMS,  CNRS-UMR  7649, Ecole Polytechnique, Route de Saclay, 91128 Palaiseau,  France}

\date{\today}
\begin{abstract}
Eukaryotic cells possess motility mechanisms allowing them not only to self-propel but also to exert forces on obstacles (to push) and to carry cargoes (to pull). To study the inherent asymmetry between active pushing and  pulling we   model   a crawling acto-myosin cell extract  as a 1D layer of active gel subjected to external forces.  We show that pushing is controlled  by protrusion and that the macroscopic signature of the protrusion dominated motility mechanism is  concavity of the force velocity relation.  Instead, pulling is driven by protrusion only at small values of the pulling force and it is replaced  by contraction when the pulling force is sufficiently large. This leads to more complex convex-concave structure of the force velocity relation, in particular, competition between protrusion and contraction can produce negative mobility in a biologically relevant range. The model illustrates active readjustment of the force generating machinery in response to changes in the dipole structure of external forces.  The possibility of switching between complementary active mechanisms implies that if necessary 'pushers' can replace 'pullers' and visa versa.
\end{abstract}
\maketitle

\section{Introduction}

Most of eukaryotic cells, including, for instance, fish keratocytes, self-propel by advancing the front and retracting the rear. A prototypical scheme of such motility includes polymerization of actin, facilitated by dynamic assembly of focal adhesions, motor-driven contraction of acto-myosin cytoskeletal network, and, finally, the detachment of adhesive contacts followed by depolymerization of actin that closes the treadmilling cycle \cite{Abercrombie1980, Dimilla1991, Stossel1993, Ridley2003, Vicente-Manzanares2005, Hoffman2009, Mogilner2009}. All three main components of the motility mechanism (polymerization, contraction and adhesion) are active and require intricate regulation as well as a continuous supply of energy. While the general crawling scheme described above is compatible with both oscillatory and steady translocation of the cell body,  in this paper we focus on steady motility modes.

The molecular and biochemical basis of cell motility is basically known, however, the qualitative understanding of the mechanical interplay between different active components is hidden behind complex computational schemes involved in modeling of cell motility \cite{Wolgemuth2005, Wolgemuth2011, Herant2010, Ziebert2012, Doubrovinski2011, Shao2010, Wang2012}. In particular, the relative mechanical role of contraction and protrusion in exerting forces on obstacles (pushing) and carrying cargoes (pulling) is usually obscured by geometrical and chemical complexity of the comprehensive mathematical models.

Protrusion is known to be the main mechanism of pushing which, for instance,  plays dominant role in Listeria propulsion \cite{John2008}. Instead, contraction is believed to be crucial for the ability of cells to pull organelles. An inherent functional disparity between protrusion-contraction components of the motility mechanism suggests a fundamental difference in the structure of the force-velocity relations associated with pushing and pulling. In experimental studies pushing and pulling are often difficult to distinguish and most of the measured force-velocity data are attributed to pushing \cite{Brunner2006, Prass2006, Schreiber2010, Zimmermann2012}.

\begin{figure}[!h]
\begin{center}
\includegraphics[scale=0.26]{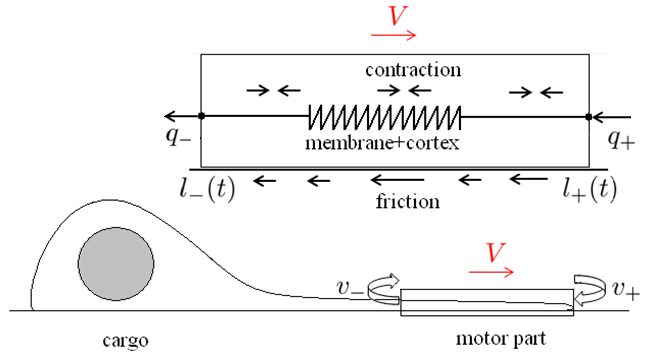}
\caption{\label{Schemecrawl} Schematic representation of an advancing lamellopodium subjected to a pushing force $q_+$ and a pulling force $q_-$.}
\end{center}
\end{figure}

To separate contributions of protrusion and contraction we use the simplest model of an active gel  and view lamellipodium  as a 1D fluid body \cite{Kruse2006}.  The actomyosin cell extract represented by such gel is assumed to be limited by free boundaries where the external loads are applied (see Fig.\ref{Schemecrawl}). Actin treadmilling also takes place on these boundaries and is modelled as an influx of mass at the front and its disappearance at the rear.  Active contraction is represented by a spatially homogeneous prestress generated at the microscale by molecular motors. Adhesion is assumed to be passive and is modeled in this minimal setting by viscous friction on a rigid background.

Our main result is that  the roles of  protrusion and contraction  may by \emph{interchangeable} depending on the character of the mechanical task performed by the cell  (pushing or pulling). We  identify an experimentally observable macroscopic signature of the dominance of each of the two mechanisms by demonstrating that pushing-dominated force-velocity relation is concave while pulling-dominated force velocity relation may be convex-concave with an interval of negative mobility.

An explicit solution of the mechanical problem shows  that in the presence of a cargo the minimal fluid model  is singular and we regularize it by accounting for an overall stiffness.  The latter may be ascribed either to a membrane or to the elastic components of the cytoskeleton \cite{Sheetz2006, Prost2007, Barnhart2010, Du2012}. Such augmentation removes singularities but preserves the main qualitative predictions of the minimal model.

To further challenge the minimal model we study the effects of spatially inhomogeneous (graded) adhesion, distributed (bulk) depolymerization and consider  the dependence of   contractile stress on  actin  density. We show that the main qualitative results regarding the convexity structure of the force velocity relation remain unchanged.

While even our augmented minimal model still underrepresents some physical effects (e.g.  active adhesion,  transport of motors, complex membrane dynamics, 3D geometry, etc.  \cite{Wolgemuth2005, Wolgemuth2011, Herant2010, Ziebert2012, Doubrovinski2011, Shao2010, Wang2012}) it allows one to go beyond force velocity relations and study the efficiency of cargo-pulling machinery.  In particular, we show that a competition between protrusion and contraction can result in a bi-modal structure of the load-efficiency relation.  By using the minimal model we could also compare the conventional kinematic mode of driving through given polymerization/depolymerization velocities with a direct control of external energy supplies responsible for protrusion and contraction.

Vis-\`{a}-vis the general  behavior of  active media, we have shown that an interplay between 'pushers' and 'pullers' \cite{Saintillan2012, Simha2002, Marchetti2012} can lead to observable effects in the presence of applied loads. The importance of the idea that different active mechanisms can swap roles depending on the task goes far beyond the subject of cell motility.

The paper is organized as follows. In Section II we formulate the minimal model, find explicit traveling wave solutions describing steadily advancing cells and study stability of these solutions. By solving the associated  transport problem we reconstruct actin density profiles in different loading regimes and reveal the mechanism behind the possibility of infinite density localization. We then study the distribution of active force dipoles in the moving cell and present an interpretation of the negative motility regime in terms of a \emph{crossover} between protrusion-dominated and contraction-dominated regimes. Elastic  regularization of the minimal model is introduced in Section III where we consider separately the  mean field (spring) model, the Kelvin-Voigt model and the Maxwell model. Other extensions of the minimal model incorporating  inhomogeneous adhesion,  distributed de-polymerization and  density dependent contraction  are analyzed in Section IV. In Section V we explore the energetics of the protrusion-contraction mechanism and study the load dependence of its efficiency. The possibility of a non-kinematic driving of the moving cell is discussed in Section VI. The final Section VII contains our conclusions.

\section{The minimal model}\label{ChapIIImod}

Our starting point is the balance of forces in a 1D layer of active gel placed on a rigid surface \cite{Kruse2006, Julicher2007}. While active dynamics of adhesion complexes is notoriously complex \cite{Bershadsky2003}, a usual assumption made in the context of cell motility  is that the time averaged tension generated by constantly engaging and disengaging focal adhesions is proportional to the velocity of the retrograde flow  \cite{Rubinstein2009, Larripa2006, Julicher2007, Shao2010, Doubrovinski2011, Hawkins2011, Tawada1991}.  If we  neglect the bi-phasic effect  \cite{Gardel2008, Gardel2010, Mogilner2009, Bois2011, Schwarz2012, Dimilla1991, Mi2007}  and assume that  the friction coefficient $\xi>0$ is constant we obtain
\begin{equation}\label{forcebalance}
\partial_x\sigma=\xi v,
\end{equation}
where $v(x,t)$ is the velocity and $\sigma(x,t)$ is the stress. Here and throughout the paper we denote by $\partial_a$ a partial derivative with respect to $a$.

By using the  constitutive model of an infinitely compressible viscous active fluid  we can write \cite{Kruse2006, Julicher2007}
\begin{equation}\label{constitutive}
\sigma=\chi+\eta\partial_x v,
\end{equation}
where $\eta>0$  is a bulk viscosity and $\chi>0$ is a constant active pre-stress. This  minimal constitutive description is clearly singular because the cell  can be infinitely stretched or  compressed and it is quite remarkable that this setting is already sufficient to capture  the essence of active competition between treadmilling and contraction.

Combining (\ref{forcebalance}) and (\ref{constitutive}) we obtain a second order differential equation which we need to solve on a domain with \emph{free} boundaries  $ l_{+} (t)$ and $l_{-} (t)$  representing the front and the rear limits of a cell. To solve this problem on a domain with \emph{fixed} boundaries we need to impose two mechanical boundary conditions
$$
\sigma(l_{\pm}(t),t)=q_{\pm}.
$$
These conditions  introduce asymmetric loading which is the central concept of this paper.  In our notations $q_+<0$ corresponds to pushing  (at the front) and $q_- >0$ to pulling (at the rear). To find the unknown functions $ l_{+} (t)$ and $l_{-} (t)$ we need to impose two additional boundary conditions. The conventional choice is a pair of kinematical constraints \cite{Kruse2006, Julicher2007}
$$
v(l_{\pm}(t),t)-\dot{l}_{\pm}=v_{\pm},
$$
where $v_{+}>0$ and $v_{-}>0$ are the polymerization and the depolymerization velocities, respectively. The prescribed sign of these velocities introduces implicit polarization of the cell which is necessary for initiation of  motility in the absence of  applied forces.

If  we now normalize length by $\sqrt{\eta/\xi}$, time by $\eta/\chi$ and stress by $\chi$, we obtain a free boundary problem which depends on four dimensionless parameters.  Two of them,  $v_{\pm}$, characterize \emph{internal} driving and the other two, $q_{\pm}$,  describe \emph{external} loading. It is natural, however, to work with a slightly different set of parameters.  Thus, parameter
$$V_m=\frac{v_-+v_+}{2}\geq 0$$
 prescribes polarity of the cell and, as we show later in the paper, gives the scale of the maximal velocity.  The remaining kinematic parameter
$$\Delta V=v_+-v_-,$$
introduces the asymmetry between polymerization and depolymerization  and, as we show later in the paper, quantifies the degree of engagement of the contractile mechanism.
 It will also be convenient to define the resultant force
$$Q=q_--q_+\geq, 0$$
which we assume to be positive and acting against the polarization direction induced by protrusion. We also introduce the force asymmetry factor
$$\epsilon=\frac{q_-+q_+}{Q},$$
which characterizes the first moment of the external force distribution. We notice that  $-1\leq\epsilon\leq 1$ with  $ \epsilon>0$ corresponding to pulling and $ \epsilon<0$ - to pushing.

The resulting dynamic problem has a peculiar structure due to an implicit assumption about separation of time scales. More specifically, the neglect of inertia means that mechanical equilibrium is reached instantaneously at the time scale of the motion of the free boundaries (Stokes flow).  The rate limiting factor is then kinetics of the free boundaries characterized by parameters $v_{\pm}$ that can be naively interpreted as describing the treadmilling process only. However, as we show later in the paper, only their sum $V_m$ can be linked to treadmilling proper while their difference $\Delta V$ is a characteristic of contraction.

\subsection{Traveling wave solutions}\label{ChapIIIfv}

The transparency of the minimal model is due to the fact that our linear force balance equation with mechanical boundary conditions can be integrated in elementary functions as  was first observed in \cite{Kruse2006, Julicher2007} for a cell without cargo. When cargo is present the velocity  profile can also be found explicitly
\begin{equation} \label{VV}
v(x,t)=\frac{A_- \cosh(l_-(t)-x)+A_+\cosh(l_+(t)-x)}{\sinh (l_+(t)-l_-(t))},
\end{equation}
where
\begin{equation} \label{VV1}
A_{\pm}=\pm(1-Q(\epsilon\pm1)/2).
\end{equation}
Knowledge of the spatial dependence and the use of kinematic boundary conditions allows one to obtain explicit equations for the functions $l_+(t)$ and $l_-(t)$. Moreover, by using the total length $L(t)=l_+(t)-l_-(t)$ we can obtain a closed dynamical problem
\begin{equation}\label{L}
\dot{L} =\Delta V+(\epsilon Q-2)\tanh\left(\frac{L}{2}\right).
\end{equation}
After this equation is solved the position of the geometrical center of the cell $G(t)=(l_+(t)+l_-(t))/2$ can be  found by  integrating a decoupled equation with the known right hand side
\begin{equation}\label{G}
\dot{G} =V_m- \frac{Q}{2\tanh (L/2)}.
\end{equation}
To specify solutions of (\ref{L}) and (\ref{G}) we need to supply the initial conditions  $L(0)$ and $G(0)$ that also fix the initial velocity profile through (\ref{VV}).

In this paper we are interested in traveling wave (TW) solutions of (\ref{L}) describing steadily translocating cells. These solutions correspond to stable  critical points of (\ref{L}) with $\dot{L}=0$ that exist if and only if
\begin{equation}\label{L11}
0<\Delta V< 2-\epsilon Q.
\end{equation}
When these conditions are satisfied the length of the cell stabilizes as $t\rightarrow\infty$ at the value
$$L_{\infty}=2\tanh^{-1}\left(\frac{\Delta V}{2-\epsilon Q}\right)>0.$$
Alongside, the function $\dot{G}$ converges to a constant $V$ given by the following force velocity relation
\begin{equation}\label{L1}
V=V_m-\frac{Q}{\Delta V}+\frac{\epsilon Q^2}{2 \Delta V}.
\end{equation}
Notice that the cell   moves to the right against the load if $V>0$ and is dragged backwards by the load if $V<0$. The maximum velocity $V^*=V_m$ is achieved when there is no load $Q=0$ and the corresponding reference length will be denoted by $L_{\infty}^*=L_{\infty}(Q=0)$.

Since the TW regimes are stable only if $2-\epsilon Q >0 $,  pushing ($\epsilon<0$) contributes to stability  while pulling ($\epsilon>0$) plays a destabilizing role. We also observe that at $\Delta V=0$ the loaded cell shrinks to a point while at $\Delta V = 2-\epsilon Q $ its length diverges. For singular solutions with $L_{\infty}=\infty$ which are only elevant in the case of pulling, the force velocity relation can be continuously extended by using (\ref{G})
\begin{equation}\label{L11}
V=V_m - Q/2.
\end{equation}
In Section \ref{Elastregu} we show that these singular solutions of the minimal model are physically meaningful and can be viewed as limits of the nonsingular solutions in the model with finite internal stiffness.

At large times  we can characterize  convergence of the initial configuration to the TW profile (transient regime)  by the formula
 $$ \vert L(t)-L_{\infty}\vert \sim e^{-t/\tau},$$
   where the characteristic time of relaxation to the steady state,  $$\tau=\frac{2(2-\epsilon Q)}{(2-\epsilon Q)^2-\Delta V^2},$$
can be measured experimentally. After this time, which depends on both, the mechanical loading and the kinematic driving  the cell can be expected to acquire the velocity predicted by the steady force-velocity relation (\ref{L1})-(\ref{L11}).

\subsection{Force velocity relation}

\begin{figure}[!h]
\begin{center}
\includegraphics[scale=0.35]{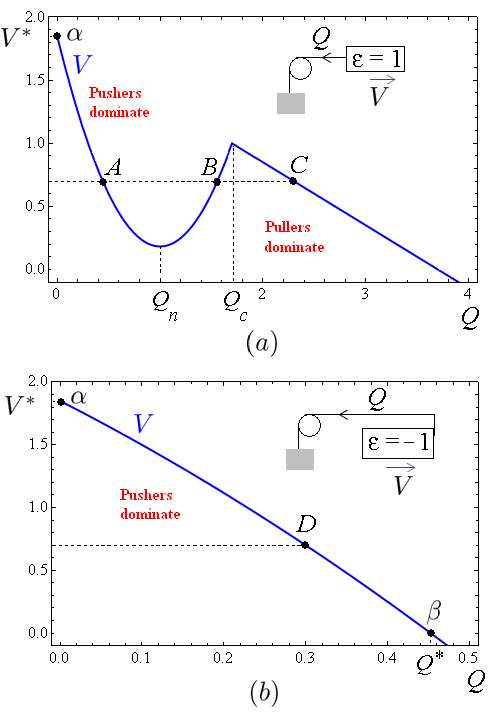}
\caption{\label{force-velocity-efficency} The typical force-velocity relations in pure pulling (a) and pushing (b) regimes. Stress, velocity and density profiles corresponding to points $A$,$B$,$C$,$D$, $\alpha$ and $\beta$  are shown in Fig.\ref{profileskinematic} and Fig.\ref{densityzeroforce}. Driving parameters are $v_-=1.7$ and $v_+=2$.}
\end{center}
\end{figure}

The structure of the obtained force-velocity relation in the ($V, Q$) plane is illustrated in Fig.\ref{force-velocity-efficency} a,b. One can see that it is markedly different for $\epsilon>0$ (pulling) and $\epsilon<0$ (pushing). The main feature distinguishing pushing from pulling is the curvature of the force velocity relation which in the regular regimes (\ref{L1}) is given by
$$\frac{\partial^2 V}{\partial Q^2}=\frac{\epsilon}{\Delta V},$$
and in the singular (pulling) regimes by
$$\frac{\partial^2 V}{\partial Q^2}=0.$$
One can see that the curvature is always negative in pushing regimes with $\epsilon<0$   which means that the corresponding force velocity  relation is concave.   Under pulling loads with  $\epsilon>0$  the force velocity curve is convex for regular regimes and is linear for singular regimes.

In the   pushing regimes the force velocity curve is characterized by the stall force $Q^{*}=(1-\sqrt{1-2\epsilon \Delta V V_m})/\epsilon$  and the maximum velocity  $V^*=V_m$, see Fig. \ref{force-velocity-efficency} (b). The concavity of the force velocity relation in this case agrees with experiments \cite{Brunner2006, Prass2006, Schreiber2010, Zimmermann2012}.
In the case of pulling, the force-velocity relation is convex for $Q<Q_c=(2-\Delta V)/\epsilon$, where  $L_{\infty}<\infty$    and is linear for $Q>Q_c$,  where  $L_{\infty}=\infty$, see Fig.\ref{force-velocity-efficency} (a). In the convex range the function $V(Q)$ is non-monotone when $\Delta V<1$ and one can distinguish two regimes:  the branch $Q<Q_n=1/\epsilon$ where the mobility is positive,  $V(Q) \sim V_m-Q/\Delta V$, and, as we show later in the paper, protrusion dominates, and the branch $Q_c>Q>Q_n$ where the mobility is negative,  $V(Q) \sim  \epsilon Q^2/(2 \Delta V)$ and the dominant active mechanism is contraction. Along the negative mobility branch the cell elongates to support larger loads till the length diverges at a critical value $Q=Q_c$. Beyond this value, we obtain configurations with infinitely separated boundary layers and mobility becomes again positive.

\subsection{Density distribution}

To interpret complex behavior of the force velocity relation in  pulling regimes we first need to reconstruct the (actin) density distribution inside the moving cell. The assumption of infinite compressibility allows one to decouple the problem of finding density distribution  from the problem of determining stress and velocity profiles.

After a 'statically determinate' mechanical problem (\ref{forcebalance}),(\ref{constitutive}) is solved, the  density $\rho(x,t)$ can be obtained from the mass transport equation:
\begin{equation}\label{kinematic}
\partial_t\rho+\partial_x(\rho v)=0,
\end{equation}
where the function  $v(x,t)$ is given by (\ref{VV}). Equation (\ref{kinematic}) must be supplemented by a single boundary condition
\begin{equation}\label{kinematic1}
 \rho(l_+(t),t)v_+ = \rho(l_-(t),t) v_-
 \end{equation}
which ensures that the exterior treadmilling mechanism  conserves  the incoming mass flux
$$\dot{m}(t)=-\rho(l_-(t),t)v_- <0.$$
Then  the total mass $M=\int_{l_-(t)}^{l_+(t)}\rho(x,t)dx,$ is constant and all actin depolymerized at the rear is instantaneously re-polymerized at the front
$$\frac{dM}{dt}=\rho(l_+(t),t)v_+-\rho(l_-(t),t)v_-=0.$$

\begin{figure}[!h]
\begin{center}
\includegraphics[scale=0.25]{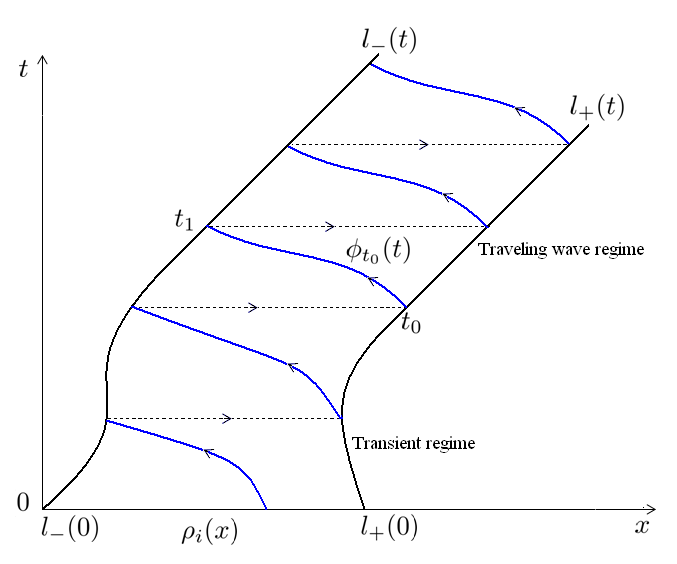}
\caption{\label{construction} Schematic structure of a particle trajectory inside a cell as it approaches the steady state TW regime.  Dotted lines indicate instantaneous treadmilling of  particles from the rear boundary of the cell to its front boundary.}
\end{center}
\end{figure}

Given an initial condition $\rho(x,0)=\rho_{i}(x)$ the mass transport problem inside the cell can be solved by method of characteristics. The initial density distribution prescribes the total mass $M$ which can be absorbed into the scaling of $\rho$ if we define dimensionless density  $\rho/\rho_0$ with $\rho_0= M/\sqrt{\eta/\xi}$. The distribution $\rho_i$ is transported in finite time along the characteristics from $[l_-(0),l_+(0)]$ to the rear boundary of the cell. The arriving mass, characterized by the   distribution $\rho(l_-(t),t)$, is then (instantaneously) transported by the treadmilling mechanism  $\text{(\ref{kinematic1})}$ from the back of the cell $l_-(t)$  to the front of the cell $l_+(t)$.  From there the mass is again transported by characteristics towards the rear boundary of the cell. This construction is then repeated indefinitely as we show in Fig.\ref{construction}.

To be more specific, consider, for instance,  a characteristic curve $x=\phi_{t_0}(t)$ originating at $l_+(t_0)$. The function  $\phi_{t_0}(t)$ is a solution of the initial value problem
\begin{equation}\label{charact}
\left\{ \begin{array}{c}
\frac{d\phi_{t_0}(t)}{dt}=v(\phi_{t_0}(t);l_-(t),l_+(t))\\
\phi_{t_0}(t_0)=l_+(t_0)
\end{array} \right.
\end{equation}
The characteristic curve $x=\phi_{t_0}(t)$  reaches the back of the cell at $t=t_1$ which can be found from the condition
 $\phi_{t_0}(t_1)=l_-(t_1).$  The density evolution along the characteristic curve can now be recovered from the transportation condition
$$\rho(\phi(t),t)=\rho(\phi(t_0),t_0)\text{e}^{-\int_{t_0}^{t}\partial_xv(\phi(u);l_-(u),l_+(u))du}.$$

\begin{figure}[!h]
\begin{center}
\includegraphics[scale=0.30]{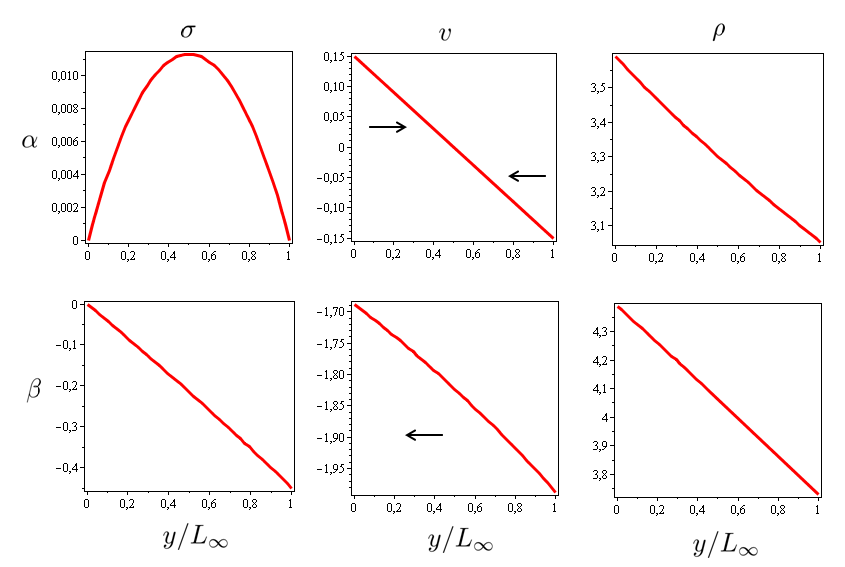}
\caption{\label{densityzeroforce} Stress, velocity and density profiles for point $\alpha$ in Fig. \ref{force-velocity-efficency} where $\epsilon=\pm1$, $Q=0$ , $V=V^*=1.85$  and for point $\beta$ in Fig. \ref{force-velocity-efficency} where $\epsilon=-1$, $Q=Q^{*}=0.45$, $V=0$.  Parameters $v_-=1.7$ and $v_+=2$. }
\end{center}
\end{figure}

In the  traveling wave regime both density and  velocity depend only on the co-moving coordinate
 $y=x-Vt,$ with $0\leq y\leq L_{\infty}$.  In particular, $\rho(x,t)=\rho(y)$. We must also have
$$
\left\{ \begin{array}{c}
l_-(t)=Vt\\
l_+(t)=L_{\infty}+Vt
\end{array} \right.
$$
Now the mass balance equation can be integrated explicitly and we obtain
\begin{equation}\label{densityfin}
\rho(y)=\frac{\dot{m}}{v(y)-V}
\end{equation}
where $\dot{m}$ is a constant mass flux  and the function $v(y)$ is given explicitly by (\ref{VV}). Since the dimensionless total mass of the cell is equal to unity, we obtain
\begin{equation}\label{massfluxTW}
\dot{m}=\left( \int_0^{L_{\infty}}\frac{dy}{v(y)-V}\right) ^{-1}.
\end{equation}
This allows us to write the final expression for the steady state density profile in the form
$$\rho(y)=\left((v(y)-V)\int_0^{L_{\infty}}\frac{du}{v(u)-V}\right)^{-1}.$$

An internal configuration of a cell at zero load ($Q=0$ , $V=V^*$), which is typical for both weak pushing and pulling, is shown in Fig.\ref{densityzeroforce} ($\alpha$).  Similar profiles for stress and velocity have been already presented in \cite{Kruse2006, Julicher2007} and here we complement the picture by presenting the associated density profile. The density accumulation at the back of the cell is in agreement with the relative velocity distribution $v(y)-V$ in the co-moving frame. One can see that this flow  is globally retrograde with higher absolute value of velocity at the leading edge than at the trailing edge.
\begin{figure}[!h]
\begin{center}
\includegraphics[scale=0.29]{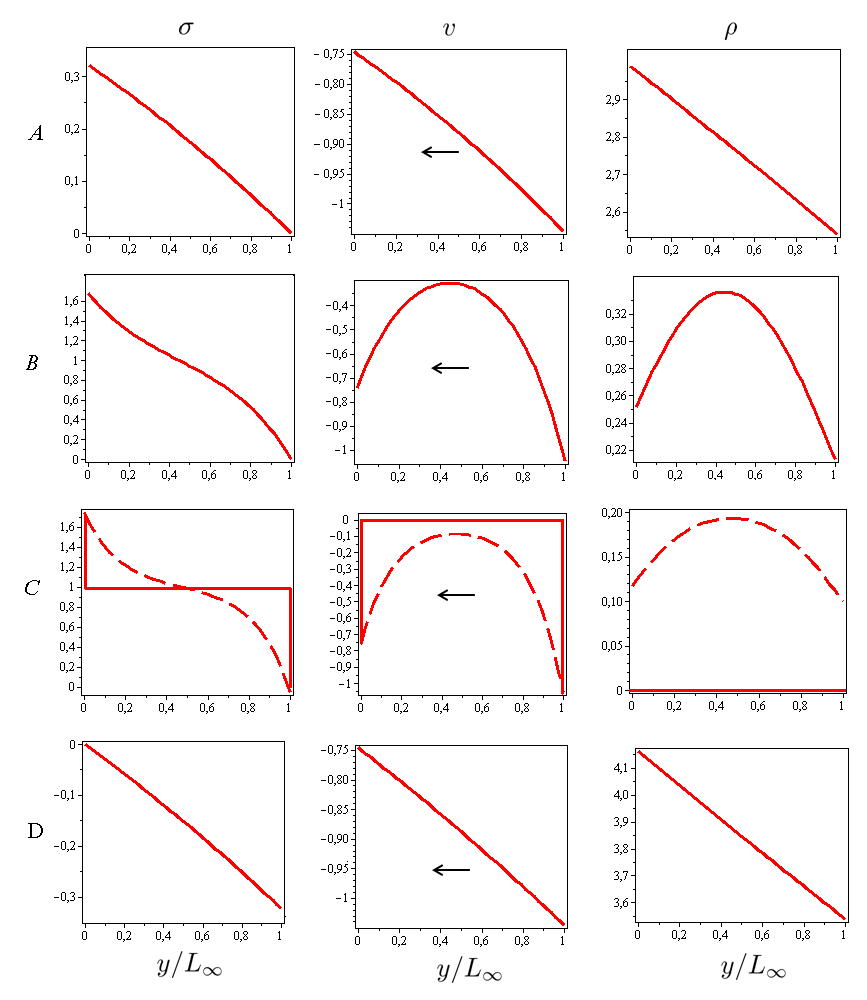}
\caption{\label{profileskinematic} Stress, velocity and density distribution inside a cell moving with the same velocity $V=0.95$ in four different loading regimes indicated as $A$, $B$, $C$ and $D$  in Fig.\ref{force-velocity-efficency}. Dashed line shows elasticity-regularized profiles corresponding to point $C'$ in Fig.\ref{fv_elast}.}
\end{center}
\end{figure}

Adding loads generates a non zero mean flow in the co-moving frame and makes the profile steeper in pushing regimes and more shallow in pulling regimes. For instance,  in Fig.\ref{densityzeroforce} ($\beta$) we show the  configuration  corresponding to (severe) pushing at the stall force conditions $Q=Q^{*}$. In this and similar regimes shown in Fig.\ref{profileskinematic} (A) and Fig.\ref{profileskinematic} (D)  the relative flow  with respect to the average velocity is prograde at the rear and  retrograde  at the front. Instead, the distribution $v(y)-V$ is globally retrograde slowing down at the rear of the cell where the density profile has a maximum, see also \cite{Kruse2006}.

A fundamentally different set of regimes, signifying, as we show in the next section, a transition from   protrusion dominated to contraction dominated motility,
is shown for the case of pulling in Fig.\ref{profileskinematic} (B) and Fig.\ref{profileskinematic} (C).  Here the relative flow with respect to the mean velocity is retrograde at both rear and front and is prograde in the central part of the cell.  The distribution $v(y)-V$ continues to be globally retrograde with strongest flow at the back and at the front of the cell which slows down in the middle part of the cell. This velocity redistribution pushes the density maximum   from the back towards the center of the cell. In Fig.\ref{profileskinematic} (C) we show that as the length of the cell diverges both stress and velocity profiles flatten everywhere outside infinitely narrow the boundary layers.

\subsection{Pushers and pullers}

In this Section we discuss  physical phenomena behind the observed differences in the structure of force velocity relations in the regimes of  pushing and pulling.

We begin with an observation \cite{Kruse2006, Carlsson2011} that the global force balance, stating that the  applied force is resisted by the friction force,
$$\int_0^{L_{\infty}}v=-Q,$$
does not distinguish between pushing and pulling. To see the role of different active agents  we need to consider the \emph{balance of couples} and by referring to asymmetric cargo we imply different signs of the dipole component of the distributed load.

By multiplying the force balance equation (\ref{forcebalance}) in the  TW regime by  $y-L_{\infty}/2$ and integrating  over the body of the cell  we obtain
$$Q\frac{\epsilon}{2}-\frac{1}{L_{\infty}}\int_{0}^{L_{\infty}}(y-\frac{L_{\infty}}{2}) v(y) dy=\frac{1}{L_{\infty}}\int_{0}^{L_{\infty}}\sigma(y)dy.$$
The first term in the left hand side
$$T_e=Q\frac{\epsilon}{2}$$
is the moment of external forces.  Since we assumed that  $Q>0$,   pulling is associated with a positive applied dipole  while pushing  - with a negative applied dipole. The  second term on the left hand side
$$T_f=-\frac{1}{L_{\infty}}\int_{0}^{L_{\infty}}(y-\frac{L_{\infty}}{2}) v(y) dy$$
represents frictional dipole which may have different signs. The integral on the right hand side defines the active dipole which can  be also rewritten as
$$T_a=\frac{1}{L_{\infty}}\int_{0}^{L_{\infty}}(1+\partial_y v)dy.$$
This term can be further decomposed into the sum $T_a=T_c+T_p$ where contraction component can be written in the TW regime as
$$T_c=1>0$$
 and protrusion  component as
 $$T_p =-\frac{\Delta V}{L_{\infty}}<0.$$
The opposite signs of these two terms suggest that the underlying active mechanisms are inherently different.  By using the terminology of the theory of active suspensions \cite{Simha2002, Saintillan2012} we can interpret the protrusion term as representing distributed "pushers"  and the contraction term  as representing distributed   "pullers".

We observe that due to the presence of a contraction (positive) force dipoles the rear boundary of the cell is pulled forward while the front boundary is pulled backward. As a compensation, contraction produces internal retrograde flow at the rear and pro-grade flow at the front. In contrast,  protrusion (negative) force dipole  pushes the rear of the cell backward while the front of the cell is pushed forward. This is compensated internally by retrograde flow at the front and pro-grade at the rear. These flows must be superimposed  with the mean flow $\bar{v}=- Q/L_{\infty}$ which is associated solely with the total applied force and is therefore always retrograde.

We can now identify  separate contributions of pushers and pullers in building the  internal configurations of the cell displayed in Fig. \ref{densityzeroforce} ($\alpha$, $\beta$) and in Fig. \ref{profileskinematic} (A-D). For instance, knowing that in the protrusion (pushers) dominated regime  the velocity gradient must be negative (retrograde at the front and pro-grade at the rear, plus a constant) we can conclude that pushers   dominate in the configurations shown in Fig. \ref{densityzeroforce} ($\alpha$, $\beta$)  and in Fig.\ref{profileskinematic} (A, D).  Similarly, if we consider regular pulling regimes exhibiting  negative mobility at $Q_c>Q>Q_n$, see  Fig.\ref{profileskinematic} (B), we can conclude that here pushers are challenged by  pullers that enforce positive velocity gradient (retrograde  at the rear and pro-grade  at the front, plus a constant).  The situation remains qualitatively similar in the singular pulling regimes illustrated in Fig.\ref{profileskinematic} (C).

Therefore, we can  identify the point $Q_n$ in Fig.\ref{force-velocity-efficency} (a) with a crossover from pushers dominated to  pullers dominated regimes. This interpretation is supported by  comparing the magnitudes of the two competing active couples.  For instance, in the realistic case $\Delta V=0.3$ \cite{Kruse2006}, illustrated in Fig.\ref{force-velocity-efficency} (a), a condition that the magnitude of the contraction couple is twice as big as the magnitude of the protrusion couple, $|T_c| \sim 2|T_p|$,  gives the value of the force $2- \Delta V/\tanh(\Delta V)=0.97$  which is quite close to the threshold $Q_n=1/ \epsilon=1$. Here it is important to mention that at $Q=0$, we have $|T_c| \sim |T_p|$ which allows the cell to eliminate frictional couple and achieve maximum velocity.

The observed crossover correlates with the transition from positive to negative mobility which also takes place at $Q_n$. Negative mobility has been discussed previously in the context of individual \cite{Cleuren2003, Ros2005, Machura2007, Haenggi2010} and interacting \cite{Brugues2009, Orlandi2010} Brownian motors. The regimes  where  velocity of the crawling cell increases with an opposing pulling force at the rear have been envisioned in \cite{Joanny2003} where negative mobility was attributed to the coupling between the velocity of retraction and the applied force $v_-(Q)$ \cite{Peskin1993}. In our model such coupling is absent which shows that negative mobility may also have a different origin.

To make quantitative predictions we use the data from \cite{Kruse2006}: $\chi = 10^3 Pa $, $\xi=5 \times 10^{16}Pa \cdot m^{-2} \cdot s$, $ \eta = 5\times 10^4 Pa \cdot s$, $v_+=2$ and $v_- = 1.7$.  This gives   for the dimensional velocity of the unloaded cell $(\chi/\sqrt{\xi \eta})V^*= 0.37\times10^{-7} m \cdot s^{-1}$ and for its dimensional length $\sqrt{\eta}{\xi}L_{\infty}^{*}=0.3\times 10^{-7} m$. This length scale is of the right order of magnitude while the velocity scale  is  at least an order of magnitude smaller than the values recorded for keratocytes and  fibroblasts \cite{Schreiber2010,Jilkine2011}. In the case of pure pushing  $\epsilon=-1$, we can use the area  $S=10^{-12}m^2$  to obtain the dimensional value of the stall force $\chi S Q^* = 1 nN $  which is realistic \cite{Brunner2006, Prass2006, Schreiber2010, Zimmermann2012}. Based on these estimates we conclude that negative motility may be expected in the interval of pulling force values  $1-1.7 nN $ and this prediction can be tested experimentally.

\subsection{Formation of singularities}

Formula (\ref{massfluxTW}) shows a possibility of the two types of degeneracies associated with reaching the condition $\dot{m}=0$. In such singular regimes the treadmilling flow becomes fully blocked.

The first type of singular behavior takes place when the length of the cell  diverges.  Here we refer to the  infinite spreading of a cell in pulling regimes with $Q>Q_c$  illustrated in Fig.\ref{profileskinematic} (C). As we show in the next Section this problem can be fixed if elastic stiffness is taken into consideration.

\begin{figure}[!h]
\begin{center}
\includegraphics[scale=0.25]{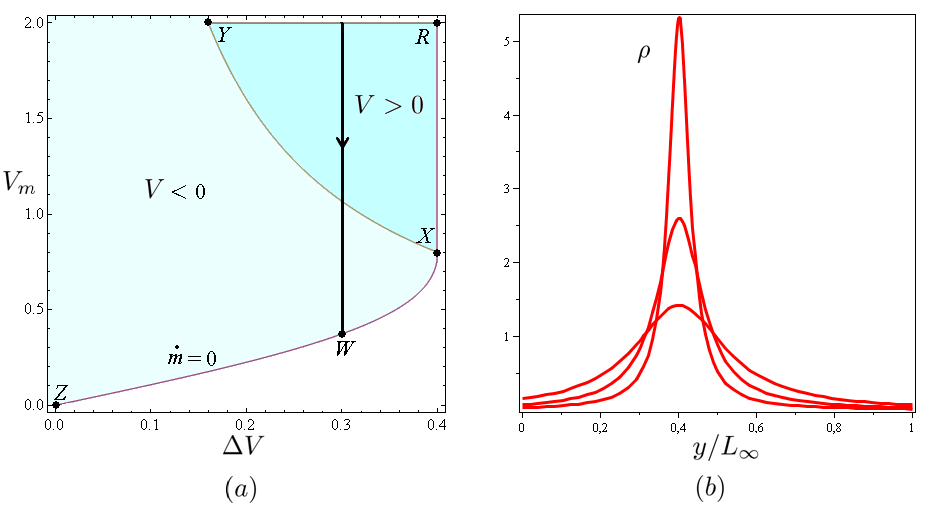}
\caption{\label{loc}  (a) Various pulling regimes in the parameter space $(V_m,\Delta V)$.  In the domain $V>0$ the cell moves against the load while if $V<0$ the cell is dragged by the load. Along the line XY the cell is static resisting the load (stall force conditions).  The singular regimes correspond to the lines ZX (infinite localization) and XR (infinite spreading). (b) Density localization along the path indicated in (a) by the solid line which ends with the formation of a singularity at point $W$. The loading is $\epsilon=1$ and $Q=1.6$. For $\Delta V=0.3$ the singularity is located at $y_1/L_{\infty}\simeq 0.4$. }
\end{center}
\end{figure}

The second  type of degeneracy is associated with non integrability of $(v(y)-V)^{-1}$ even for cells with finite lengths. Such singularity can take place when there exists a point $y_0$ where $v(y_0)=V$.  To illustrate the possibility of this type of singular behavior consider the whole set of pure pulling regimes shown in Fig.\ref{loc} (a).   Notice that the line $\dot{m}=0$ in the ($V_m,\Delta V$) plane consists of two segments: XR and ZX. The segment XR is associated with infinite spreading of the cell as discussed above. Instead, along the segment ZX the length of the cell remains finite while the density localizes infinitely in a single  point inside the cell.
\begin{figure}[!h]
\begin{center}
\includegraphics[scale=0.21]{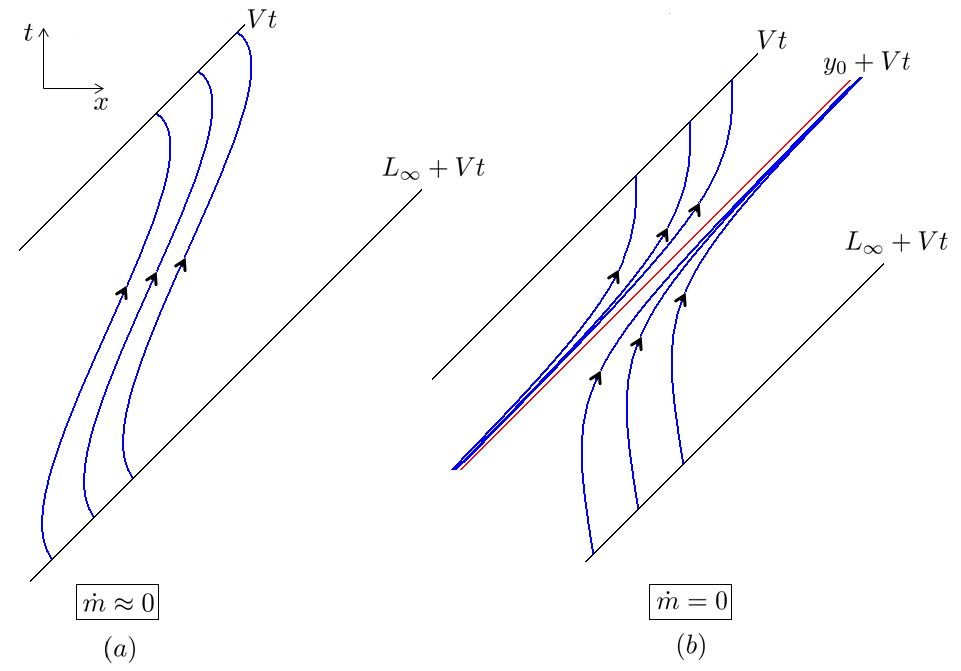}
\caption{\label{zeroflux} Sketch of particles trajectories as the steady state (TW) mass flux approaches zero,  see Fig.\ref{loc} for the related density profiles. (a) At small values of $\dot{m}$ each particle spends considerable  in  a small region near a line $x=y_0+Vt$. (b) When $\dot{m}=0$ all particle trajectories converge to the line $x=y_0+Vt$ which leads to a blow up singularity.}
\end{center}
\end{figure}

To locate this point consider a regular ($\dot{m}\neq 0$) density  profile with a local maximum at $y=y_1$ where $\partial_y\rho(y_1)=0$. Such point can be found from the equation
$$A_- \sinh(y_1)=A_+ \sinh(L_{\infty} -y_1).$$
One can see that $y_1$ does not depend on $V_m$ and if we  lower the value of $V_m$ till the  regime with $\dot{m}= 0$ is reached (see the solid trajectory ending at W in Fig.\ref{loc} (a)) we obtain $y_0=y_1$. The associated phenomenon of infinite mass concentration at the point $y=y_1=y_0$  is illustrated in Fig.\ref{loc} (b).
Below the line (ZWX) the reconstruction of mass density is not possible because of the interpenetration of matter.

The phenomenon of infinite density localization can be also illustrated through the behavior of the characteristics (particle trajectories). In Fig.\ref{loc} we show how characteristics in the TW regime concentrate as one approaches point W in Fig.\ref{loc} (a). Even before reaching the regime W the particles spend considerable time around the line $x=y_0+Vt$, see Fig.\ref{loc} (a). At the point W where $\dot{m}=0$  the mass flow gets completely blocked as we show in Fig.\ref{loc} (b). Notice also that due to decoupling of mechanical and mass transport problems in the minimal problem, the velocity field  in such singular regimes remains regular.
Similar to the case of infinite spreading, the problem of infinite localization can be resolved if we take into consideration internal stiffness of the cell body.

\section{Elastic regularization}\label{Elastregu}

A natural way to regularize the minimal model is to introduce an intermediate-time stiffness of the cell. Such stiffness prevents the unloaded cells from contraction-induced collapse and sets the rest length and it also keeps this length from diverging in the case of pulling.

Elasticity may be associated either with the cytoskeleton or with the cell membrane.  Membrane and cortex elasticity can be modeled  in a prototypical setting as a mean field  elastic feedback provided by elastic springs linking different parts of the cell \cite{Barnhart2010, Du2012}. Visco-elastic properties of cytoskeleton strongly depend on the characteristic time of the problem \cite{Boal2002, Chen2010, Mofrad2009} and the corresponding corrections to the active gel model in the bulk of the cell are usually incorporated either in the framework of a short time (Maxwell) elastic model \cite{Julicher2007,Callan-Jones2011, Joanny2007, Ranft2012, Rubinstein2009} or a long time  (Kelvin-Voigt) elastic model \cite{Pathak2008, Larripa2006, Banerjee2011}.

\subsection{Mean field elasticity}

The simplest elastic regularization of the minimal model is through mean field coupling between the leading and trailing edges of a cell \cite{Sheetz2006, Prost2007, Barnhart2010, Du2012}. If this coupling is linear elastic, the applied loads become
$$q_{\pm} \rightarrow q_{\pm}+ k\frac{L-L_0}{L_0},$$
where $k>0$ is a dimensionless stiffness and $L_0$ is a prescribed dimensionless reference length. The   meaning of parameter $L_0$ is clear from the fact that for $k>1$ and $V_m=\Delta V=0$  there exists a nontrivial static solution with $L_{\infty}=L_0(1- 1/k)$ (preferred shape).

\begin{figure}[!h]
\begin{center}
\includegraphics[scale=0.38]{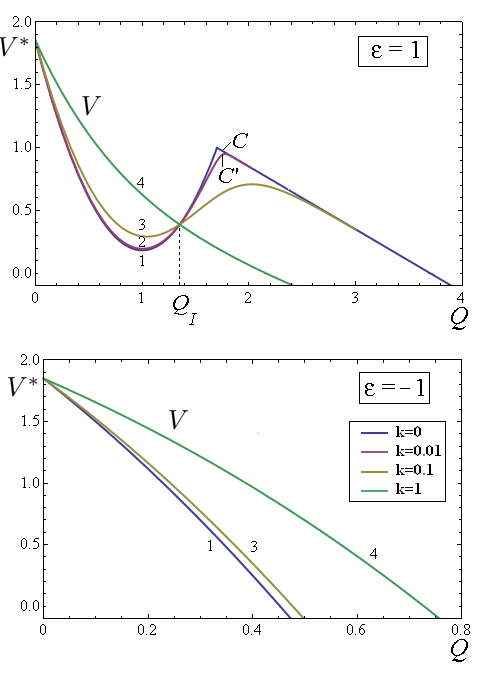}
\caption{\label{fv_elast}  Force velocity relations in pure pushing and pulling modes with different  $k_{1,2,3,4}=\left\lbrace 0,0.01,0.1,1\right\rbrace $ and $L_0=1$. Driving parameters are $v_-=1.7$ and $v_+=2$. Internal profiles corresponding to points $C$ and $C'$ are presented in Fig.\ref{profileskinematic} (C). The minimal model is recovered at $k=0$}
\end{center}
\end{figure}

In dynamics the steady state (TW) solution is now stable for all $\Delta V>0$ and to find $L_{\infty}(Q)$ one needs to solve
$$
\Delta V=(2-\epsilon Q +2k\frac{L_\infty-L_0}{L_0})\tanh\left( \frac{L_\infty}{2}\right) .
$$
Then, the force velocity relation can be found from the relation
$$V(Q)=V_m- \frac{Q}{2\tanh\left(\frac{L_\infty (Q)}{2}\right)}
$$
and its $k$ dependence is illustrated in Fig.\ref{fv_elast}. We observe that independently of the value of $k$ all force-velocity curves cross at $Q=0$ where $V=V^*$.  The second common intersection point at
$$Q_I=\frac{1}{\epsilon}\left( 2-\frac{\Delta V}{\tanh\left(\frac{L_0}{2}\right)}\right)$$
exists when $\epsilon>0$ and $L_0>L_{\infty}^*$.

\begin{figure}[!h]
\begin{center}
\includegraphics[scale=0.28]{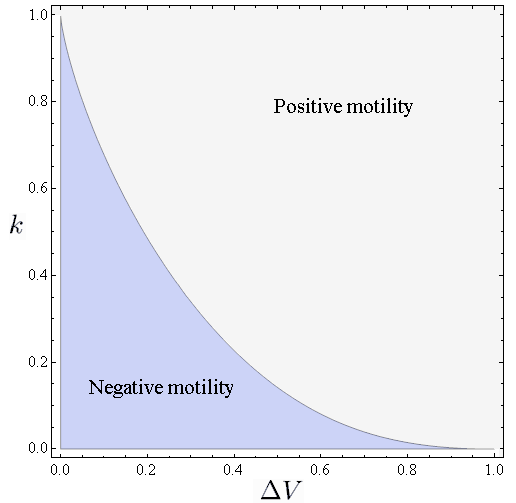}
\caption{\label{NegMotElast} Domain of negative mobility in the parameter space $(k,\Delta V)$. The boundary between regimes with positive and negative mobility is given by the function $k=k^*(\Delta V)$.}
\end{center}
\end{figure}

The salient feature of the regularized model is that at  $k \rightarrow 0$ the mean field  force velocity curves approach their minimal model counterparts  including both the regular regimes with finite cell lengths and the singular regimes with infinite cell lengths. However, despite similarity in shape between the force velocity curves  in the minimal model and in the regularized model with $k \sim 0$,  the  length of the cell in the regularized model is always \emph{finite} so that infinite stretching, undermining the minimal model, does not take place. Unfortunately,  by using the mean field elasticity model one cannot also remove the localization singularity in the minimal model because in this regularized setting the force balance remains independent from the mass balance.

The phenomenon of negative mobility for the pulled cells survives in the mean field model and disappears only at a critical value  of the stiffness $k=k^*(\Delta V)$, see Fig. \ref{NegMotElast}. The qualitative difference in convexity between pulling and pushing persists beyond $k^*(\Delta V)$, see for instance regime with $k=1$ in Fig.\ref{fv_elast}. However, at  $k\gg k^*(\Delta V)$ the force-velocity relations associated with pushing  and  pulling regimes become similar.

The available data on static configurations \cite{Sheetz2006, Prost2007, Barnhart2010, Du2012} suggest that the dimensionless parameter $k$ (normalized by $\chi$) must be in the range $1-10$ which  apparently excludes the negative mobility regimes. However, many models of cell dynamics are built under the assumption that  long time elasticity is negligible and essentially assume that $k=0$ \cite{Kruse2006, Rubinstein2009, Callan-Jones2011, Ranft2012}. The ambiguity maybe due to the ability of cytoskeleton  to fluidize by engaging active cross-linkers  that can modify its stiffness over at least two orders of magnitude \cite{Koenderink2009, MacKintosh2010, Sheinman2012}.

In view of such broad rheological flexibility of the cytoskeleton, the effective stiffness may easily reach below the threshold $k=1$ which means that negative motility regimes cannot be excluded in vivo and can be probably artificially engineered in vitro through partial suppression of the  stiffening components of acto-myosin network. If we use numerical values of parameters from Section \ref{ChapIIIfv} we find that $\chi S Q_I\simeq 1.3 nN$ which gives the scale of pulling forces where  negative mobility can be expected. Notice that this value is above the resolution of an atomic force microscopy cantilever which has been previously used in the measurements of force velocity curves \cite{Prass2006}.

\subsection{Kelvin-Voigt elasticity}

Instead of using spring based elastic regularization considered above we can directly incorporate distributed  elasticity into the constitutive model. The closest to the mean field  model is the Kelvin-Voigt model accounting for the elastic response at long time scales. In the 1D setting we need to assume that
$$
 \sigma=\chi+\eta\partial_xv-p(\rho),
$$
where $p(\rho)$ is the stress-density relation. In our version of Kelvin-Voigt model we further assume that this relation is linear 
$$p(\rho)=E \left(\frac{\rho}{\rho_r}-1\right),$$
where $\rho_r$ is the reference density and $E$ the elastic modulus \cite{Boal2002, Mofrad2009}. The resulting system of coupled non-dimensional equations can be written as 
$$
\left\{ \begin{array}{c}
\partial_t\rho+\partial_x(\rho \partial_x\sigma)=0\\
-\partial_{xx}\sigma+\sigma=1-K (\rho/\hat{\rho}_r -1),
\end{array} \right.
$$
where we introduced two new nondimensional parameters $K= E/\chi\text{ and } \hat{\rho}_r = \rho_r/\rho_0 $. To find the steady state (TW) regimes we need to solve a simpler system
 $$
\left\{ \begin{array}{c}
-\partial_{yy}\sigma+\sigma=1-K(\frac{\rho}{\hat{\rho}_r}-1)\\
\rho(y)=\left[ \left(\partial_y\sigma(y)-V\right)\int_0^{L_{\infty}}\frac{du}{\partial_u\sigma(u)-V}\right]^{-1}
\end{array} \right.
$$
with mechanical boundary conditions,  
 $\sigma(0)=q_-$ and $\sigma(L_{\infty})=q_+,$
and  kinematic boundary conditions,  
 $V=v_-+\partial_y\sigma(0)$  and $V=v_++\partial_y\sigma(L_{\infty}).$

\begin{figure}[!h]
\begin{center}
\includegraphics[scale=0.28]{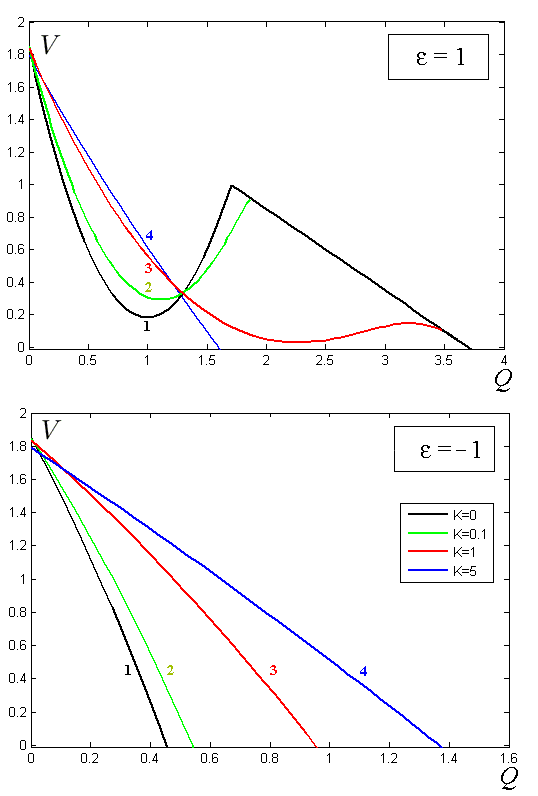}
\caption{\label{fv_Kelvin} Force-velocity curves for  the Kelvin-Voigt model with different $K_{1,2,3,4}=\left\lbrace 0,0.1,1,5\right\rbrace $.  Other parameters are $\hat{\rho}_r=1$, $v_-=1.7$ and $v_+=2$. The minimal model is recovered at $K=0$.}
\end{center}
\end{figure}

The ensuing force velocity relations are shown in Fig.\ref{fv_Kelvin} for different values of $K$. Qualitatively, these curves are quite similar to their analogs in the   mean field model, in particular, the negative mobility regimes persist for sufficiently small $K$. However, the problem with the divergence of the cell length at finite $Q$ does not disappear which means that Kelvin-Voigt regularization of tension is weaker than in the mean field model. The reason is that the linear stress density dependence in the bulk does not penalize sufficiently the infinite stretching of the gel layer. 

We observe, however,  that in the framework of linear Kelvin-Voigt elasticity the density singularities shown in Fig.\ref{loc} (b) disappear, which means that this model regularizes infinite compression adequately, something the mean field model could not accomplish. This suggests that  the Kelvin-Voigt model and the mean field models show complimentary features and should be used in combination.

\subsection{Maxwell elasticity}

In contrast to two elastic regularization schemes considered above,  Maxwell model associates elasticity with fast time scales. In the interpretation of this model with co-rotational (Jaummann) convective derivative \cite{Julicher2007, Callan-Jones2011, Joanny2007} the dimensional problem can be written as
\begin{equation}\label{Max}
\left\{ \begin{array}{c}
\partial_t\rho+\partial_x(\rho v)=0\\
\xi v=\partial_x\sigma\\
(\eta/E) \partial_t\sigma+v\partial_x\sigma +\sigma=\chi+\eta\partial_x v
\end{array} \right.
\end{equation}
where $E$ is the (infinite frequency) elastic modulus. One can see that in this setting the mechanical problem decouples again from the mass transport problem.

\begin{figure}[!h]
\begin{center}
\includegraphics[scale=0.28]{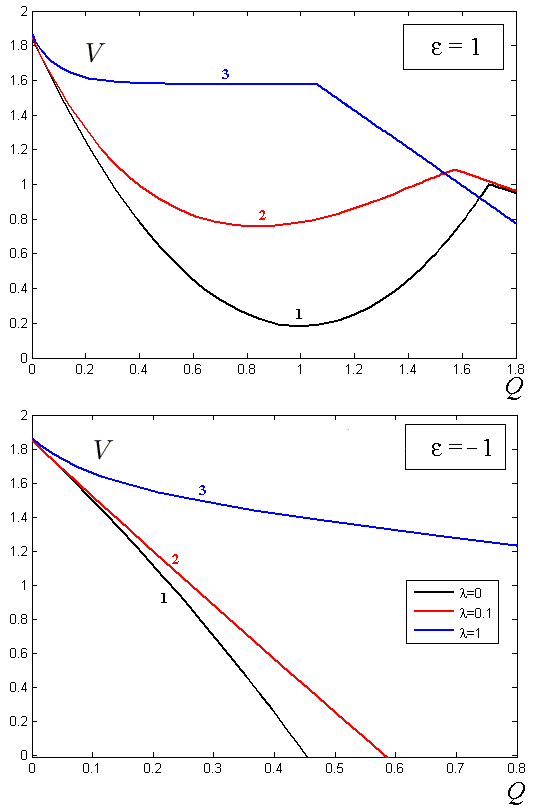}
\caption{\label{fv_Maxwell}  Force-velocity  curves for the Maxwell model with different $\lambda_{1,2,3}=\left\lbrace 0,0.1,1\right\rbrace $. Other parameters are $\rho_0=1$, $v_-=1.7$ and $v_+=2$. The minimal model is recovered at $\lambda=0$.}
\end{center}
\end{figure}

A single dimensionless equation describing steady state (TW) regimes takes the form
\begin{equation}\label{Max1}
\lambda\partial_y\sigma(\partial_y\sigma-V)-\partial_{yy}\sigma+\sigma=1
\end{equation}
where the new non dimensional parameter is $\lambda= \eta/E$ and experimental data suggest that $\lambda=0.02-0.2$ \cite{Rubinstein2009, Wottawah2005, Kole2005, Panorchan2006, Mofrad2009}. The equation (\ref{Max1}), which is  nonlinear in contrast to what we have had in the minimal model, must be again supplemented by two  mechanical boundary conditions,
 $\sigma(0)=q_-$ and
$\sigma(L_{\infty})=q_+,$
and two kinematic boundary conditions,
$ \partial_y\sigma(0)-V=-v_-$ and
$\partial_y\sigma(L_{\infty})-V=-v_+.$

The nonlinear boundary value problem (\ref{Max1}) was  studied numerically and in Fig.\ref{fv_Maxwell} we show the force velocity relations corresponding to different values of  $\lambda$.   One can see that the negative motility regimes survive at finite $\lambda$ which suggests that the qualitative behavior observed in the minimal model is stable under this regularization.  

Introducing Maxwell elasticity, however, fails to regularize the infinite stretching singularity. Moreover, Maxwell model does not allow for static equilibria describing the rest state of a cell and there is a numerical evidence that the problem with infinite localization of mass also persists. We can then conclude that at least in the study of steady motility regimes the combination of Kelvin-Voigt and mean field elasticity should be preferred to the use of Maxwell elastic regularization.

\section{Miscellaneous}

To check robustness of our predictions we study in this Section three different extensions of the minimal model \emph{not dealing} with elasticity.
The first extended model allows for inhomogeneous friction, in the second  model depolymerization is assumed to be taking place everywhere in the bulk of the cell body and in the third  active contractile pre-stress becomes a function of actin density.

\subsection{Inhomogeneous friction}

Assume that in the steadily moving cell the friction coefficient $\xi$ is graded from rear to front. For instance, $\xi$ may be viewed as  proportional to the steady state density of focal contacts which are known to concentrate in the frontal part of the advancing lamellipodium  \cite{Bottino2002,Zajac2008,Rubinstein2009}.

More specifically, suppose that  $\xi=\xi \kappa(z)$ where $z=y-L_{\infty}/2$ and $-L_{\infty}/2\leq z\leq L_{\infty}/2 .$  The dimensionless mechanical equations describing TW regimes takes the form
\begin{equation}\label{TWadh}
-\partial_z\left( \frac{\partial_{z}\sigma}{\kappa(z)}\right) +\sigma=1.
\end{equation}
While this equation is still linear, it now has a variable coefficient. The mechanical boundary conditions remain the same as in the minimal model
 $\sigma(-L_{\infty}/2)=q_-$ and $\sigma(L_{\infty}/2)=q_+$,
but the kinematic boundary conditions get modified 
$$V=\frac{\partial_z\sigma}{\kappa}(-L_{\infty}/2)+v_-=\frac{\partial_z\sigma}{\kappa}(L_{\infty}/2)+v_+.$$

A semi-explicit solution of the resulting Sturm-Liouville problem can be expressed in terms of two linearly independent functions $A(z)$ and $B(z)$ solving the following elementary sub-problems \cite{Mikhlin1960}:
$$
\left\{ \begin{array}{c}
A''=\kappa(z) A\\
A'(-L_{\infty}/2)=1=A'(L_{\infty}/2),
\end{array} \right.
$$
and
$$
\left\{ \begin{array}{c}
B''=\kappa(z) B\\
B'(-L_{\infty}/2)=1\text{ and }B'(L_{\infty}/2)=-1.
\end{array} \right.
$$
By using the functions $A(z)$ and $B(z)$, we can write the  force velocity relation in the following explicit form
$$
\left\{ \begin{array}{c}
\Delta V=(2-Q\epsilon)[A]+Q\{A\}\\
V=V_m+\frac{2-Q\epsilon}{2}[B]+\frac{Q}{2}\{B\},
\end{array} \right.
$$
where $ 2[f]= f(L_{\infty}/2)-f(-L_{\infty}/2) $  and  $ 2\{f\}= f(L_{\infty}/2)+f(-L_{\infty}/2)$ .

Suppose, for instance, that $\kappa(z)=1+\theta \kappa_1(z)$ ,  where $\theta$ is a small parameter and the function $\kappa_1(z)$  is odd. Then, in the lowest order in  $\theta$ we obtain:
$$
\left\{ \begin{array}{c}
[A]=\{B\}^{-1}=-\tanh(L_{\infty}/2)\\
\{A\}=-[B]=-\frac{\theta}{2\sinh(L_{\infty})}\int_{-L_{\infty}/2}^{L_{\infty}/2}\sinh(2z)\kappa_1(z)dz
\end{array} \right.
$$
The resulting force velocity relation can be written semi-explicitly
\begin{equation}\label{KK1}
\left\{ \begin{array}{c}  
\Delta V=(2-Q\epsilon)\tanh(\frac{L_{\infty}}{2})- \\
\frac{2Q\theta}{\sinh(L_{\infty})}\int_{0}^{L_{\infty}/2}\sinh(2z)\kappa_1(z)dz\\
V=V_m+\frac{\theta(2-Q\epsilon)}{\sinh(L_{\infty})}\int_{0}^{L_{\infty}/2}\sinh(2z)\kappa_1(z)dz-
\\ \frac{Q}{2\tanh(\frac{L_{\infty}}{2})}
\end{array} \right.
\end{equation}

Observe that if  the integral $\int_{0}^{L_{\infty}/2}\sinh(2z)\kappa_1(z)dz$ in (\ref{KK1}) is positive which means that if there is a frictional bias at the front,  the cell will have larger length and will move with larger velocity than in the minimal model with $\theta=0$. If, instead, the friction is stronger at the back, the cell will have smaller length and will move slower than in the minimal model. These results are compatible with the observation that adhesion complexes predominantly position themselves at the front of the moving cell \cite{Bottino2002, Zajac2008, Rubinstein2009} which can then be interpreted as an optimization of velocity.
\begin{figure}[!h]
\begin{center}
\includegraphics[scale=0.33]{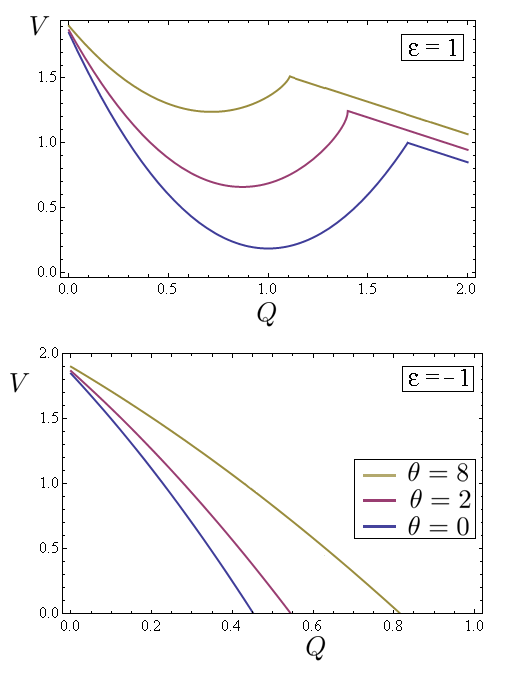}
\caption{\label{adhforce} Force velocity relations in the case of inhomogeneous friction with $\theta$ as a parameter.The minimal model is recovered at $\theta=0$.}
\end{center}
\end{figure}

In Fig.\ref{adhforce} we show numerical results for finite values of $\theta$. To ensure that  the  concentration of adhesive complexes at the front is four times larger than in the back, which is  plausible for keratocytes \cite{Rubinstein2009}, we must take  $\theta=8$. From Fig.\ref{adhforce} we see that at this level of inhomogeneity the general shape of the force velocity curves remains qualitatively the same as in the minimal model.

\subsection{Delocalized depolymerization}

We recall that in the minimal model the mass transport problem is fully decoupled from the force balance problem. As a result the  density distribution does not affect the force velocity relation and different models of actin transport can be made compatible with the same force velocity relation.

To exploit this idea we consider in this Section  a version of the minimal model where depolymerization is not localized at the rear front of the cell. Instead,  we assume that depolymerization takes place everywhere in the bulk of the cell as it is suggested by observations \cite{Theriot1991, Svitkina1997, Keren2009, Vallotton2005}.

As a first step we exclude localized depolymerization  by putting $v_-=0$.  Then we modify the mass conservation equation by adding a source term. If we make the simplest assumption that  the rate of depolymerization is a linear function of density we obtain
\begin{equation} \label{dens}
 \partial_t\rho+\partial_x(\rho v)=-\hat{\gamma} \rho.
\end{equation}
The coefficient $\hat{\gamma}$ can be estimated in the range $0.01-0.05s^{-1}$ \cite{Larripa2006, Rubinstein2009, Barnhart2011}.

The proposed reformulation of the minimal problem affects the velocity distribution only  through the specific choice of one of the kinematic fluxes (condition  $v_-=0$)  which gives $\Delta V=v_+$ and $V_m=v_+/2$. The  explicit solution (\ref{VV}) and the general formulas (\ref{L}) remain valid. The stability condition takes the form
$$0<v_+<2-\epsilon Q.$$
One can see that the assumption  $v_+\simeq 2$, which we used throughout the paper to illustrate the results obtained in the minimal model,  is no longer adequate in the case of pulling. We therefore assume a smaller value $v_+\simeq 1$ which is  also plausible in view of \cite{Svitkina1997, Larripa2006, Keren2009, Julicher2007}. In the presence of elasticity penalizing infinite stretching  such re-scaling is not necessary.

The treadmilling boundary condition  for (\ref{dens}) can now be written in the form
\begin{equation}
\rho(l_+(t),t)=\frac{\hat{\gamma} M}{v_+}.
\end{equation}
It ensures that the total mass remains constant
$$\frac{dM}{dt}=-\hat{\gamma} M+\rho(l_+(t),t)v_+=0.$$
We can again absorb $M$ into the scaling of $\rho$ by using dimensionless variable  $\rho/\rho_0$. The ensuing nondimensional problem depends on the new parameter $\gamma = \eta \hat{\gamma}/\chi$ which can be estimated in the range  $\sim 0.5-2.5$.

The dimensionless equation describing the TW regimes takes the form
\begin{equation} \label{KK2}
\partial_y(\rho(v-V))=-\gamma \rho
\end{equation}
where we recall that $y=x-Vt$. If we now introduce the treadmilling mass flux $\dot{m}=-\gamma$,  we may  write the solution of (\ref{KK2}) explicitly
\begin{equation}\label{densityexp}
\rho(y)=\frac{\dot{m}}{v(y)-V}\exp\left(-\int_y^{L_{\infty}}\frac{\dot{m}}{v(u)-V}du\right).
\end{equation}
Here the pre-exponential factor is exactly the same as in the minimal model (\ref{densityfin}) while the new exponential term describes modulation due to distributed depolymerization.
\begin{figure}[!h]
\begin{center}
\includegraphics[scale=0.39]{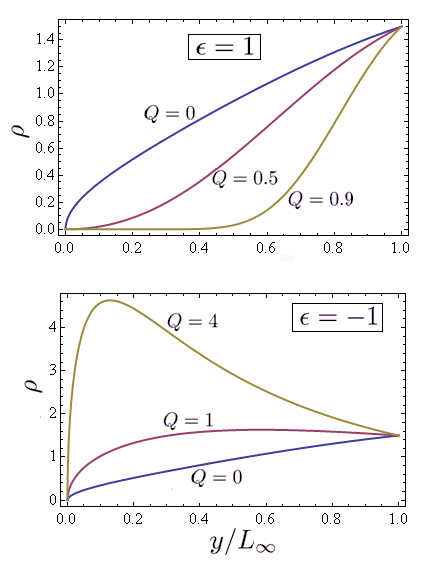}
\caption{\label{newdensity}  Density distribution in the pulling and pushing regimes for the model with delocalized depolymerization. Parameters: $v_+=1$ and $\gamma=1.5$ }
\end{center}
\end{figure}

To illustrate the role of this term  we now show that in this new setting we can obtain a peak of density at the front of the cell and a decay in the back which is the pattern typically observed in moving cells \cite{Svitkina1997, Keren2009, Schreiber2010}. To this end we  explicitly compute the values of actin  density  at the front $\rho(L_{\infty})$ and at the back $\rho(0)$. The first of these quantities can be found directly
$$\rho(L_{\infty})=\frac{\gamma}{v_+}.$$
To find the second quantity  we need to use an asymptotic development of the integral term in (\ref{densityexp}) at  small $y$
$$\rho(y)\sim \frac{1}{A_+}\frac{A_--A_+}{-2A_+\sinh(L_{\infty})}y^{\frac{\gamma}{A_+}-1},$$
where $A_+$ and $A_-$ are defined in (\ref{VV1}). From this formula we see that in the relevant range  $\gamma>A_+$  (the assumption $\gamma>1$ ensures this inequality)  we obtain $$\rho(0)=0.$$ Notice that now we have  $\rho(0)<\rho(L_{\infty})$ while in the minimal model we always had $\rho(0)>\rho(L_{\infty})$. 

In Fig.\ref{newdensity} we choose $\gamma=1.5$ and $v_+=1$ and show the typical density profiles for both pure pushing and pure pulling cases. One can see that even for unloaded cells the distributed depolymerization leads to a global decay of actin density from the front to the back. Interestingly, this decay may not be monotone.  It is clear,  however, that the fine structure of the density at the front of the cell is dominated by microscopic interactions of the actin network with the membrane and cannot be captured by our simplified model.

\subsection{Density dependent contraction}

The fact that contractile pre-stress  depends on both actin and myosin densities is well established \cite{Alt1995, Alt1999, Stephanou2008, Bois2011, Recho2013}.  Therefore the assumption that the corresponding active term is equal to a constant is one of the main weaknesses of the minimal model.

A rather general theory of  active gels implying coupling between the active prestress and the transport of different components of  actomyosin network has been developed in \cite{Kruse2000, Kruse2001, Kruse2003a}.  While it was shown that this coupling induces a rich variety of dynamic behaviors, the problem was typically studied in a fixed domain and without external loading. In particular,  the issue of the force velocity relation was not addressed. 

In this Section we study  force velocity relations under the assumption that  contractile stresses depends on cytoskeletal density representing actin filaments with the same orientation.  At the same time we neglect the important coupling of active stress with motor density studied in \cite{Bois2011, Hawkins2011, Recho2013}.

We  begin by writing the  system of coupled equations of the model  where it is convenient to distinguishing three subproblems:  

Force balance:
\begin{equation}\label{force_balance}
\left\{ \begin{array}{c}
-\partial_{xx}\sigma+\sigma=\sigma_a(\rho)\\
\sigma(l_-(t),t)=q_- \text{ and } \sigma(l_+(t),t)=q_+
\end{array} \right.
\end{equation}
Polymerization/depolymerization:
\begin{equation}\label{growth model}
\left\{ \begin{array}{c}
\dot{l}_-=v_-+\partial_x\sigma(l_-(t),t)\\
\dot{l}_+=v_++\partial_x\sigma(l_+(t),t)\\
l_-(0)=l_-^0<l_+(0)=l_+^0
\end{array} \right.
\end{equation}
Actin transport:
\begin{equation}\label{actin_conservation}
\left\{ \begin{array}{c}
\partial_t\rho+\partial_x(\rho\partial_x\sigma)=0\\
\rho(l_-(t),t)v_-=\rho(l_+(t),t)v_+ \\
\rho(x,0)=\rho_i(x).
\end{array} \right.
\end{equation}
The  force velocity relation can be obtained by solving these equations numerically for different loading conditions and tracing the solutions till they approach  different traveling wave regimes.
\begin{figure}[!h]
\begin{center}
\includegraphics[scale=0.31]{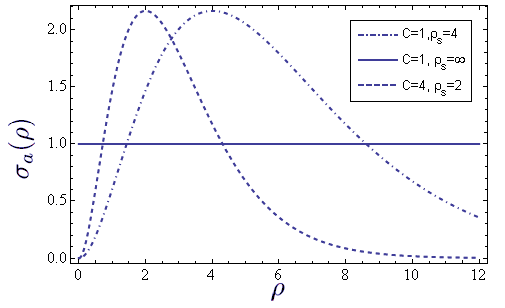}
\caption{\label{nonlinearcontractility} Contractile stress $\sigma_a$ as a function of actin density $\rho$ for different choices of parameters $\rho_s$ and $C$ taken from \cite{George2012}. Minimal model is recovered at $C=1$, $\rho_s=\infty$. }
\end{center}
\end{figure}

In our numerical experiments we observed that by assuming a linear dependence of active prestress on actin density $\sigma_a(\rho)$ we do not reach TW regimes and obtain instead oscillatory modes of cell motility \cite{Kruse2003a}. The situation changes if we assume  that this dependence is non-monotone which agrees with microscopically motivated models considered in \cite{Alt1995, Alt1999, Stephanou2008, George2012}. According to these models at small actin densities   more   filaments allow more motors to bound and to induce contractile stresses, however, there is a density threshold after which compaction of the network prevents further increase of the contractile stresses. By accepting this reasoning we  used the  function $\sigma_a(\rho)$   proposed in \cite{George2012} which in dimensionless form can be written as
\begin{equation}\label{force_balance1}
\sigma_a(\rho)= C\rho^2 \exp(-2\frac{\rho}{\rho_s}).
\end{equation}
Here $C$ is a constant and $\rho_s$ is the actin saturation density, see Fig.\ref{nonlinearcontractility}. 

In our numerical experiments  the initial location of the cell boundaries was at $l_-^0=0$, $l_+^0=1$. The initial density distribution was chosen to be $$\rho_i(x)=\frac{2}{1+v_-/v_+}\left(1+\left(\frac{v_-}{v_+}-1 \right)x\right), $$
which is the simplest way to satisfy the boundary conditions and the requirement that $\int_{l_-^0}^{l_+^0}\rho=1$. By varying the initial data we could reach different traveling wave  regimes and in this way recover the full force-velocity relation.

\begin{figure}[!h]
\begin{center}
\includegraphics[scale=0.29]{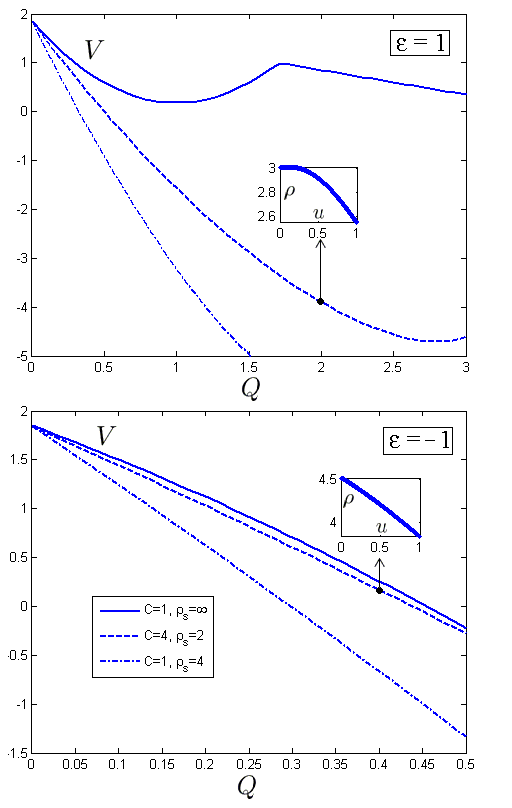}
\caption{\label{coupledactin}  Typical force velocity relations for the model with actin dependant contractility. Parameters are $v_-=1.7$, $v_+=2$.  Inserts show the steady states density profiles located  on the decreasing limb  of the density-contractility curve, see Fig.\ref{nonlinearcontractility}. Minimal model is recovered at $C=1$, $\rho_s=\infty$.}
\end{center}
\end{figure}

The results are presented in Fig.\ref{coupledactin}. As we see,  the imposed coupling does not destroy the fundamental difference in convexity properties between the force velocity curves in pushing and pulling regimes.  We also observe that in the parametric regimes presented in Fig.\ref{coupledactin} the steady density profiles always lie on the decreasing limb  of the density-contractility curve shown in Fig.\ref{nonlinearcontractility}. Further studies are needed to understand this phenomenon as well as other effects including oscillatory and stick slip type non equilibrium steady states \cite{Kruse2000, Kruse2001, Kruse2003a}.
 
\section{The efficiency of cargo transport}

The simplicity of the minimal model allows one not only to obtain explicit force velocity relations but also to study the  energetics of a self propelling  cell carrying a cargo. 

If we multiply the force balance equation (\ref{forcebalance}) by $v(x,t)$ and use the constitutive realtion (\ref{constitutive})  we obtain the global energy balance equation
$$\int_{l_-(t)}^{l_+(t)} v^2+\int_{l_-(t)}^{l_+(t)}(\partial_xv)^2+\int_{l_-(t)}^{l_+(t)} \partial_x v=[\sigma v]_{l_-(t)}^{l_+(t)}. $$
In this equation we can identify the following terms
\begin{enumerate}
\item 
$D_{f} = \int_{l_-(t)}^{l_+(t)} v^2 >0,$  dissipation rate associated with surface friction.
\item $D_v= \int_{l_-(t)}^{l_+(t)}(\partial_xv)^2>0,$  dissipation rate associated with bulk viscosity.
\item $P_{c}=-\int_{l_-(t)}^{l_+(t)}\partial_x v>0$, rate of energy consumption by the contractile mechanism.
\item $P_p=(qv)_--(qv)_+>0$, rate of energy consumption by the protrusion mechanism.
\item $A=(q\dot{l})_--(q\dot{l})_+$, the power expanded against the external forces.
\end{enumerate}
In the case of TW regimes all these terms can be computed explicitly. In particular, by using nondimensional quantities we obtain,
 $$P_{c} =\Delta V, P_p= Q V_m-\frac{\epsilon Q\Delta V}{2}, A =QV.$$
In these notations we can write the energy balance in the form
\begin{equation} \label{eff11}
P_p+P_c=A+D,
\end{equation}
where $D=D_{f}+  D_v.$ The  mechanical efficiency of cargo transportation can then be defined as follows:
\begin{equation} \label{eff}
\Lambda= W/H.
\end{equation}
Here the numerator $W$ describes the useful work  and may in addition to $A$ contain an additive  Stokes term $P_S=LV^2$, which is nonzero even in the absence of the cargo \cite{Lighthill1952, Wang2002, Suzuki2003}. However in our problem this correction can be shown to be small and will be neglected.

The denominator
 $H>0$ describes external energy supply associated with ATP hydrolysis which drives the motility process. 
It is clear that $H$ must include the power $H_A=P_p+P_c$ exerted by active forces on the constraining environment. It should also contain the 'maintenance' term  $H_D$ which accounts for energy consumption required to sustain the active state in the absence of macroscopic motion \cite{Hill1938}. By using terminology introduced in \cite{Julicher1997, Julicher2007} for weakly non-equilibrium regimes we can identify  $H_A$ and  $H_A$ with the terms that are, respectively,  linear and quadratic in the measure of chemical non-equilibrium $\Delta\mu$. In what follows, we neglect the 'quadratic' term $H_D$ (dealing with degrees of freedom that are invisible in our macro-scopic model)  comparing to the 'linear' term $H_A$.
\begin{figure}[!h]
\begin{center}
\includegraphics[scale=0.29]{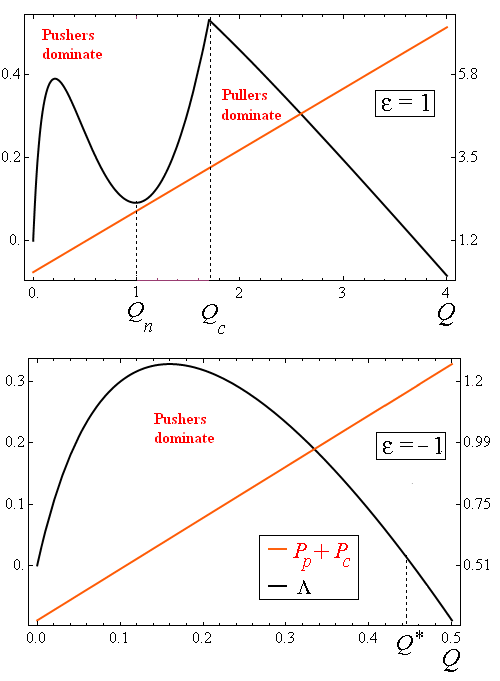}
\caption{\label{efficencyIII} Rate of energy consumption and efficiency as functions of the load in the cases of pure pulling and pushing loading modes. The corresponding force velocity relation is shown in Fig.\ref{force-velocity-efficency}. Driving parameters are $v_-=1.7$ and $v_+=2$.}
\end{center}
\end{figure}

By using these simplifying assumptions we can write
\begin{equation} \label{eff11}
\Lambda= \frac{A}{P_p+P_c}.
\end{equation}
By using the fact that $D>0$ one can showthat $0<|\Lambda|<1$. In the TW limit the efficiency (\ref{eff11}) can be computed explicitly
$$
\Lambda=\frac{Q V}{\Delta V+QV_m-\epsilon Q \Delta V/2}.
$$

In Fig.\ref{efficencyIII} we show  the efficiency $\Lambda$ and the energy consumption  rate $P_p+P_c$ as functions of the total load $Q$ for the TW regimes presented in Fig.\ref{profileskinematic}.  First of all we  observe that the divergence of the cell length in the minimal model in the pulling regimes with $Q\geq Q_c$ does not lead to singular behavior of any of the energetic measures. Second, we notice, that while in the case of pushing the function $\Lambda(Q)$ displays a usual single maximum,  in the case of pulling the efficiency-load relation  becomes \emph{bi-modal}. The two maxima can be identified with protrusion dominated and contraction dominated motility mechanisms. Such bi-modality may carry biological advantages allowing a the cell to switch back and forth between two highly efficient regimes by controlling, for instance, the friction experienced by the nucleus.

\begin{figure}[!h]
\begin{center}
\includegraphics[scale=0.31]{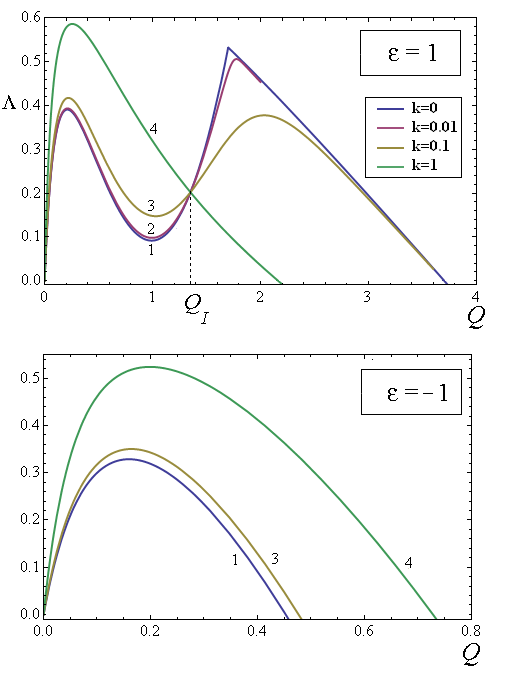}
\caption{\label{efficencyelast} Efficiency as a function of the load in the elasticity-regularized model in pure pushing and pulling regimes with different  $k_{1,2,3,4}=\left\lbrace 0,0.01,0.1,1\right\rbrace $ and $L_0=1$. Experimental data suggest that $k=1-10$ (see \cite{Barnhart2010, Du2012}). Parameters: $v_-=1.7$ and $v_+=2$. The corresponding force velocity relations are shown in Fig.\ref{fv_elast}. Minimal model is recovered at $k=0$.}
\end{center}
\end{figure}

If we augment the minimal model by adding mean field elasticity (see Section \ref{Elastregu}) the energy balance equation takes the form
$$P_p+P_c+P_e=A+D,$$
where the new term $P_e$ describes the power exerted by protrusion mechanism against the elastic 'spring' 
$$P_e=-k\frac{L-L_0}{L_0}(\dot{L}-\Delta V). $$
This term remains nonzero in  TW regimes where $\dot{L}=0$ and it 
should be added to the denominator in the expression of the efficiency
\begin{equation} \label{eff1}
\Lambda= \frac{A}{P_p+P_c+P_e}.
\end{equation}
In Fig.\ref{efficencyelast} we show the $k$ dependence of the efficiency (\ref{eff1}). One can see that the two peak structure of the function $\Lambda(Q)$ survives in the regularized model till a  threshold in $k$, signifying also the disappearance of the negative mobility range,  is reached.

\section{Alternative driving modes}\label{Alternat}

\begin{figure}[!h]
\begin{center}
\includegraphics[scale=0.19]{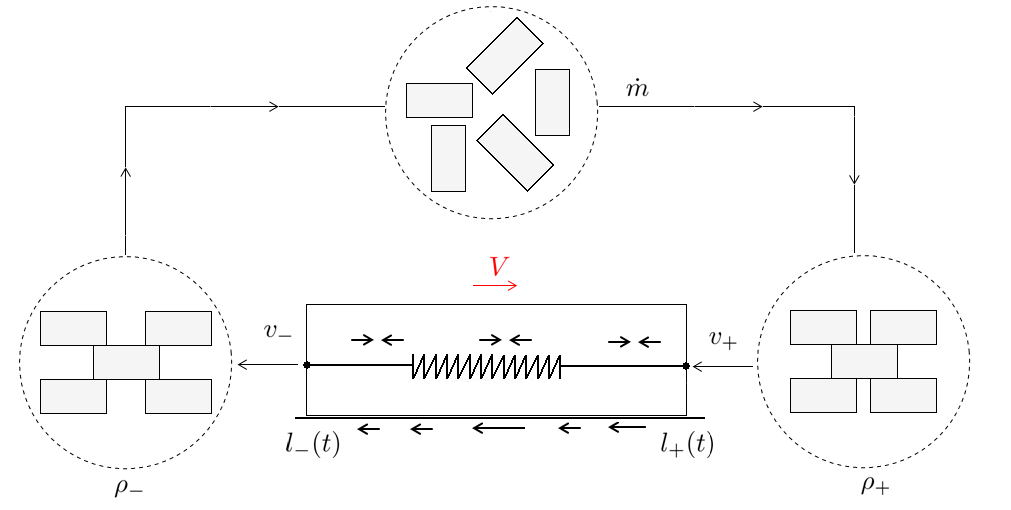}
\caption{\label{treadcycle} Schematic structure of the treadmilling cycle showing different densities of arriving  (polymerizing) and departing (depolymerizing) material.}
\end{center}
\end{figure}

In the minimal model we used an assumption that the process of cell motility is driven by the kinematic fluxes characterized by parameters $v_+$  and $v_-$. This assumption, illustrated in Fig. \ref{treadcycle},  means that we impose separately the \emph{velocities} of polymerizing (arriving) and depolymerizing (departing) mass points, see also \cite{Kruse2006, Larripa2006, Julicher2007, Mogilner2009, Rubinstein2009}. The fact that nothing is said about  the  \emph{densities} of the arriving or departing material allows one to decouple the mechanical problem from the mass transportation problem and makes the analysis fully transparent. This transparency, however, comes at a cost.

First, it is clear that the treadmilling is characterized by only one parameter, the mass flux $\dot{m}$, so by fixing two parameters $V_m$ and $\Delta V$ we are implicitly constraining both treadmilling and contraction. This is also clear from the fact that parameter $\Delta V = v_+-v_-$ serves as a measure of energy consumption in the contraction mechanism. Second, by prescribing the kinematic fluxes  $v_+$ and $v_-$  we have no direct control of the treadmilling mass flux.  As a result we encounter singular regimes with $\dot{m}=0$ which leads to either infinite mass localization inside the cell or to infinite spreading of the cell body. Third, by focussing on kinematic fluxes we do not put any restrictions on the energy consumption required to sustain different active mechanisms which appears to be a natural biological constraint.

Notice also that the problem setting where driving is performed through parameters $v_+$ and $v_-$ contains an implicit assumption that the  material arrives with a particular density (particular structural organization).  Another implicit assumption is that the departing material has a density which depends on the activity of the contractile machinery.  While these assumptions are plausible, they may not be the most natural ones from the biological point of view.  Even more importantly, as we have shown in the previous Sections, these assumptions necessarily lead to singularities.

In this Section, to further challenge the robustness of our conclusions about the asymmetry between pushing and pulling, we consider an alternative modality of driving  by imposing constraints on energetic rather than kinematic parameters.  The main difficulty in dealing with non kinematic driving schemes is that they couple the mechanical and the mass transport problems already in the minimal setting.

\begin{figure}[!h]
\begin{center}
\includegraphics[scale=0.23]{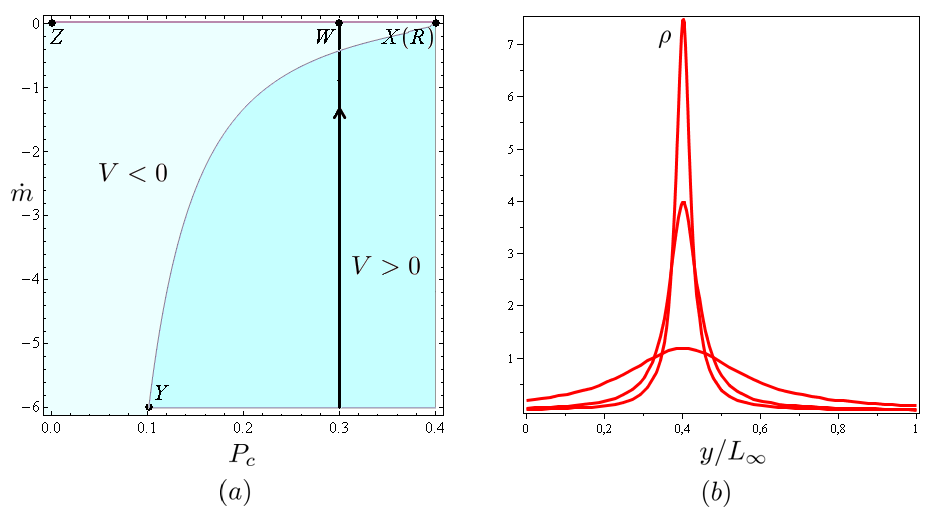}
\caption{\label{loc2} (a) Various pulling regimes in the parameter space $(\dot{m},P_c)$.  The singular regimes correspond to the lines ZX (infinite localization) and to the point X(R) (infinite spreading). (b) Density localization along the path indicated in (a) by the solid line which ends with the formation of a singularity at point $W$. The loading is $\epsilon=1$ and $Q=1.6$. Analogue of Fig.\ref{loc}. }
\end{center}
\end{figure}

More specifically, we assume that the cell controls the treadmilling rate, characterized by the total mass flux $\dot{m}<0$, and the energetics of the contraction process, characterized by the consumed power $P_c=\Delta V$. The advantage of this new parametrization is that protrusion and contraction can now be controlled independently. If we choose the pair $(\dot{m},P_c)$ as the parameters instead of ($V_m, \Delta V$), we again obtain stable TV solutions given that $P_c<2-Q\epsilon \text{ and } \dot{m}<0$. The proposed driving mode is in fact equivalent to the kinematic driving mode in the TW regimes because the Jacobian of the transformation $(v_-,v_+) \rightarrow (P_c((v_-,v_+)),\dot{m}(v_-,v_+))$
$$det \left(\begin{array}{cc}
\frac{\partial P_c}{\partial v_-}&\frac{\partial \dot{m}}{\partial v_-}\\
\frac{\partial P_c}{\partial v_+}&\frac{\partial \dot{m}}{\partial v_+}\\
\end{array} \right)  =\frac{\int_0^{L_{\infty}} \frac{dy}{(v(y)-V)^2}}{(\int_0^{L_{\infty}} \frac{dy}{v(y)-V})^2}\geq \frac{1}{L_{\infty}}>0 $$ 
is strictly positive for $0<L_{\infty}<\infty$.

By using the parametrization $(\dot{m},P_c)$  we can easily avoid the density localization phenomenon illustrated in Fig.\ref{loc} without introducing elasticity. To illustrate this point we show  in  Fig.\ref{loc2} the pulling TW  regimes in the parameter plane $(\dot{m},P_c)$  where we again distinguish between regimes where cell carries the cargo ($V>0$) and regimes where it is dragged by the cargo ($V<0$). Fig.\ref{loc2} has to be compared with Fig.\ref{loc} where the same regimes are shown in the  $(V_m,\Delta V)$ space; the only difference is that now the line $XR$ corresponding to regimes with $L_{\infty}=\infty$ collapses on a single point $X(R)$. It is clear that if the treadmilling flux is prescribed so that $\dot{m}\neq 0$, the singularities associated with the line $ZX(R)$ in Fig.\ref{loc} are automatically excluded.

\begin{figure}[!h]
\begin{center}
\includegraphics[scale=0.27]{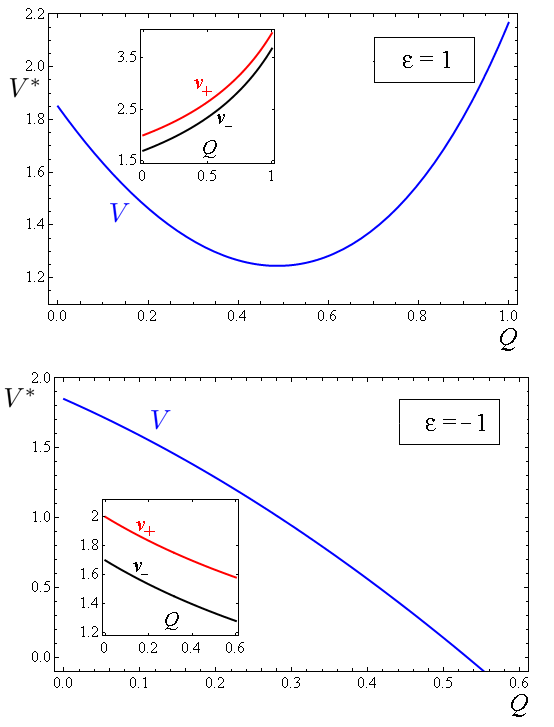}
\caption{\label{push_pull_ener} Force velocity relations in pure pushing and pulling TW regimes when driving is performed by imposing $\dot{m}=-6.1$ and $P_c=0.3$. Inserts show the ensuing dependances of $v_+$ and $v_-$ on $Q$. }
\end{center}
\end{figure} 

In Fig. \ref{push_pull_ener} we show the force velocity relations in the minimal model with  prescribed $(\dot{m},P_c)$. One can see that the qualitative difference between pushing and pulling endures in this new setting, moreover, we again observe regimes with negative mobility. It is interesting that by fixing  parameters $P_c$ and $\dot{m}$ we induce a dependence of the polymerization and depolymerization rates ($v_-,v_+$) on $Q$ (see the inserts in Fig. \ref{push_pull_ener}) which agrees qualitatively with the trends suggested in \cite{Kruse2006} based on the polymerization ratchet model. We also note that at sufficiently strong pulling loads $Q>Q_c=(2-P_c)/\epsilon$, the cell length $L_{\infty}$ diverges which suggests that also in the case of non-kinematic driving the minimal model should still be elastically regularized.

Finally we remark that instead of the pair $(\dot{m},P_c)$ we could also prescribe another set of energy related parameters, for instance, $(P_p,P_c)$. Ultimately, the choice of the driving mode requires an understanding of the microscopic side of the model and the answer may depend on the type of the cell, the environment and the regime of loading.

\section{Conclusions}

In this paper we used the simplest  model of a crawling cell to study an interplay between contraction and protrusion required to sustain and carry various cargoes. The model describes a layer of active gel subjected to external forces. It extends previous studies focussed predominantly on the behavior of unloaded active media or on problems with fixed boundaries.

By using an analytically transparent framework provided by the minimal model we demonstrated for the first time that contraction and protrusion mechanisms can interchange their roles as one varies the dipole component of the external load. Our model predicts a possibility of a relatively sharp transition between protrusion dominated motility and contraction dominated motility in response to an increase of the pulling force.  This transition  has a macroscopic signature and can be in principle identified experimentally with a negative mobility range on a force-velocity curve.

The advantage of the minimal setting is that it delivers explicit steady state solutions describing asymmetrically loaded self-propelling cells and allows analytical access to their stability.  Only in such prototypical framework the competition between contraction and treadmilling can be studied in the transparent form without any geometric effects obscuring the interplay between competing active ingredients of the model. The simplicity of the model allowed us to elucidate  active adjustment of the force producing machinery to the subtle changes in the character of external loading. The possibility of such adjustment implies that 'pushers'  both collaborate and compete with 'pullers'.

The augmentation of the active gel model involving elastic stresses in addition to viscous and active stresses was found to be essential for the removal of singularities inherent in the minimal model. In particular,  mean field elasticity appears to be the most universal way of introducing a resting configuration even when treadmilling is absent and to deal with infinite spreading at finite pulling loads. Instead, Kelvin-Voigt visco-elastic model while also ensuring  existence of static configurations and removing infinite density localization, fails to secure the finite length of the cell in the whole interval of applied pulling loads. We conjecture that a combination of  mean field and   Kelvin-Voigt elastic terms in the system of equations describing active gels is sufficient to fully regularize the minimal model.
 
To make definitive predictions about the feasibility of the  negative mobility regimes, focussed measurements of the effective stiffness associated with different elastic structures of the cell are necessary. In the situation when elastic coupling strongly affects the force velocity relations,  studying kinetic relations for differently loaded cells may be the way to furnish a set of independent bounds on such stiffness. It should be emphasized, however,  that our conclusions regarding convexity-concavity structure of the force velocity relation are much less sensitive to the value of the stiffness than the very existence of the negative motility regime. Thus, pushing and pulling force velocity relations remain qualitatively different even when the negative motility regime disappears.

Perhaps our most intriguing finding is that the fine structure of the force-velocity relation may depend on the modality of external driving and we argued that kinematic driving may not be the only physically and biologically natural choice. In particular, we suggested that instead of the rates of polymerization and depolymerization, the cell may be controlling the energy supplies required for the functioning of contraction and protrusion mechanisms.  We have shown, however, that while the detailed shape of the force velocity relation depends on the choice of the driving mode, its loading-sensitive convexity-concavity structure is a robust feature of the model.

\section{Acknowledgements}

The authors thank J.F. Joanny,  K. Kruse, A. Mogilner and two anonymous reviewers for helpful comments. The work of P.R. was supported by the Monge Doctoral Fellowship from Ecole Polytechnique

\end{document}